\documentstyle[a4,11pt]{article} \begin{document} \begin{titlepage}
\vskip 2cm
\begin{flushright}
Preprint CNLP-1997-04
\end{flushright}
\vskip 2cm
\begin{center}
{\bf
Soliton equations in N-dimensions
as exact reductions of Self-dual
Yang - Mills  equation IV. The (2+1)-dimensional  mM-LXII and Bogomolny
equations}
\footnote{Preprint
CNLP-1997-04. Alma-Ata. 1997 }
\end{center}
\vskip 2cm
\begin{center}
Kur. Myrzakul and R.  Myrzakulov
\footnote{E-mail: cnlpmyra@satsun.sci.kz}
\end{center}
\vskip 1cm

\begin{center}
 Centre for Nonlinear Problems, PO Box 30, 480035, Alma-Ata-35, Kazakhstan
\end{center}

\begin{abstract}
Some aspects of the multidimensional soliton geometry are considered.
The relation between soliton equations in 2+1 dimensions and the Self-Dual
Yang-Mills and Bogomolny equations are discussed.
\end{abstract}


\end{titlepage}

\setcounter{page}{1}
\newpage

\tableofcontents

\Large
\section{Introduction}

\quad 1985. R. S. Ward: {\it "many (and perhaps all?) of the ordinary and partial differential
equations that are regarded as being integrable or solvable
may be obtained from the self-dual gauge field equations
(or its generalizations) by reductions"} [1].

1991. M. J. Ablowitz and P. A. Clarkson: {\it "This strongly suggests that
the  KP and DS equations can be obtained from SDYM by exact reductions
(i.e., no asymptotic limits)"} [1].

1996. R.M. : {\it "Many soliton equations in 2+1 dimensions, such as, the
DS, KP, mKP, KdV, mKdV  ( and so on ) equations are exact reductions
of the M-LXII or mM-LXII equations. At the same time, their spin
equivalent counterparts such as  the Ishimori, M-X, M-I, M-IX
( and so on ) equations are exact reductions of the  mM-LXI or M-0
equations"} [20].

1997. At present,  it is well known  that the SDYM equation (SDYME)
is integrable [2,23] and
contains as reductions many  soliton equations
(SEs) [3-7, 23-28, 35]. Moreover from the SDYME can obtains may be the more
known representatives of (2+1)-SEs -the
KP and DS equations in asymptotic limit [1].

So some SEs are exact reductions of the SDYME, at the same time, some SEs
are exact reductions of the "absolutely" other equations (from the soliton
geometry), e.g.,
of the mM-LXII equation.
In this note,
we try understand this situation, that is, first aim of this paper.

Also we try understand some aspects of the multidimensional soliton
geometry, that is, second aim of this note (on the soliton geometry
see e.g. [8-20, 36-38]).

\section{Equations of surfaces in 1+1 dimensions}

In this section we review briefly the some known basic facts
from 1+1 dimensions to set our notations and terminology.
We consider a surface  evolving in 3-dimensional space.
We denote local coordinates of the surfaces by $u^{1}=x, u^{2}=t$.
The surface is specified by the position vector ${\bf r}(u^{1}, u^{2})
=(x^{1}, x^{2}, x^{3})$.
Let the metric, i.e., the first fundamental form of this surface,
be given by
$$
I=d{\bf r}^{2}=g_{\alpha\beta}du^{\alpha}u^{\beta}  \eqno(1)
$$
As usual, the extrinsic curvature, i.e., the second fundamental
form is defined as
$$
II=-d{\bf r}d{\bf n}=b_{\alpha\beta}du^{\alpha}u^{\beta}  \eqno(2)
$$
In (1)-(2)
$$
g_{\alpha\beta}={\bf r}_{\alpha}{\bf r}_{\beta}, \quad
b_{\alpha\beta}={\bf r}_{\alpha\beta}\dot {\bf n},
\quad {\bf r}_{\alpha}=\frac{\partial {\bf r}}{\partial u^{\alpha}} \eqno(3)
$$
${\bf n}=(n^{1}, n^{2}, n^{3})$ is the unit normal vector and
$$
n^{i} =
(det g)^{-\frac{1}{2}}\epsilon^{ikm}\frac{\partial x^{k}}{\partial u^{1}}
\frac{\partial x^{m}}{\partial u^{2}}  \eqno(4)
$$
Here $\epsilon^{ikm}$ is a totally antisymmetric tensor
with $\epsilon^{123}=1$.   Repeated indices are summed
on unless otherwise noted.
The Gauss-Weingarden equation (GWE) reads
$$
{\bf r}_{\alpha\beta}=\Gamma^{\gamma}_{\alpha\beta}{\bf r}_{\gamma}+
b_{\alpha\beta}{\bf n} \eqno(5a)
$$
$$
{\bf n}_{\alpha}=-g^{\gamma\beta}b_{\alpha\gamma}{\bf r}_{\beta}  \eqno(5b)
$$
Here, the Christoffel symbols $\Gamma^{\gamma}_{\alpha\beta}$ of the
second kind  defined as usual
$$
\Gamma^{\gamma}_{\alpha\beta}=
\frac{1}{2}g^{\gamma\lambda}(\frac{\partial g^{\lambda\beta}}
{\partial u^{\alpha}}+
\frac{\partial g^{\alpha\lambda}}{\partial u^{\beta}}-
\frac{\partial g^{\alpha\beta}}{\partial u^{\lambda}})\eqno(6)
$$
From the compatibility conditions of (5), we get
the Gauss-Codazzi-Mainardi-Peterson equation (GCMPE)
$$
R_{\gamma\alpha\beta\lambda}=
b_{\gamma\beta}b_{\alpha\lambda}-b_{\gamma\lambda}b_{\alpha\beta}
\eqno(7a)
$$
$$
D_{\mu}b_{\alpha\beta}=D_{\alpha}b_{\mu\beta}
\eqno(7b)
$$
In the above,
$$
R^{\gamma}_{\alpha\beta\lambda}=
\frac{\partial \Gamma^{\gamma}_{\alpha\lambda}}{\partial u^{\beta}}
-\frac{\partial \Gamma^{\gamma}_{\alpha\beta}}{\partial u^{\lambda}}
+ \Gamma^{\gamma}_{\mu\beta} \Gamma^{\mu}_{\alpha\lambda}
- \Gamma^{\gamma}_{\mu\lambda}\Gamma^{\mu}_{\alpha\beta}
\eqno(8)
$$
is the Riemann tensor and
$$
D_{\mu}f_{\alpha}=\frac{\partial }{\partial u^{\mu}}f_{\alpha}-
\Gamma^{\gamma}_{\alpha\mu}f_{\gamma}, \quad
D_{\mu}f^{\alpha}=\frac{\partial }{\partial u^{\mu}}f^{\alpha}+
\Gamma^{\alpha}_{\lambda\mu}f^{\lambda} \eqno(9)
$$
are the covariant derivatives. The Gaussian curvature $K$ and
the mean curvature $H$ of
the surface are
$$
K=det(g^{\mu\nu}b_{\nu\lambda})=\frac{1}{g}R_{1212}, \quad g=det(g_{ij})  \eqno(10)
$$
$$
H=tr(g^{\mu\nu}b_{\nu\lambda})=\frac{1}{2}g^{\mu\nu}b_{\mu\nu}
\eqno(11)
$$

Among the global characteristics of surfaces we mention the integral curvature
$$
\chi = \frac{1}{2\pi}\int_{S}Kg^{\frac{1}{2}}d^{2}u \eqno(12)
$$
where $K$ is the Gaussian curvature and integration in (12) is performed
over the surface. For compact oriented surfaces
$$
\chi=2(1-q) \eqno(13)
$$
where $q$ is the genus of the surface and we will generally assume
that surfaces are compact and oriented unless otherwise specified.

Sometimes, it is convenient to work in orthogonal basis
$$
{\bf e}_{1}=\frac{1}{g_{11}}{\bf r}_{u^{1}}, \quad {\bf e}_{2}={\bf n},
\quad {\bf e}_{3}={\bf e}_{1}\wedge{\bf e}_{2} \eqno(14)
$$
Then the GWE takes the form
$$
\left ( \begin{array}{ccc}
{\bf e}_{1} \\
{\bf e}_{2} \\
{\bf e}_{3}
\end{array} \right)_{x}= C
\left ( \begin{array}{ccc}
{\bf e}_{1} \\
{\bf e}_{2} \\
{\bf e}_{3}
\end{array} \right)
\eqno(15a)
$$
$$
\left ( \begin{array}{ccc}
{\bf e}_{1} \\
{\bf e}_{2} \\
{\bf e}_{3}
\end{array} \right)_{t}= G
\left ( \begin{array}{ccc}
{\bf e}_{1} \\
{\bf e}_{2} \\
{\bf e}_{3}
\end{array} \right) \eqno(15b)
$$
where
$$
C =
\left ( \begin{array}{ccc}
0             & k     & -\sigma \\
-\beta k      & 0     & \tau  \\
\beta \sigma  & -\tau & 0
\end{array} \right) ,\quad
G =
\left ( \begin{array}{ccc}
0       & \omega_{3}  & -\omega_{2} \\
-\beta\omega_{3} & 0      & \omega_{1} \\
\beta\omega_{2}  & -\omega_{1} & 0
\end{array} \right) \eqno(16)
$$
and $k, \tau, \sigma$ are some functions of
$g_{\alpha\beta}, b_{\alpha\beta}$. Here ${\bf e}_{j}$
is the moving trihedral of a  surface and
$$
{\bf e}_{1}^{2}=\beta = \pm 1, \quad {\bf e}_{2}^{2} = {\bf e}^{2}_{3} = 1
\eqno(17)
$$
$k, \sigma$ and $\tau$ are called the normal curvature, geodesic curvature,
and geodesic torsion, respectevily. So GCMPE takes the form
$$
C_{t} - G_{x} + [C, G] = 0 \eqno(18)
$$

We note that for  the curves case, the  equation (15a) is as $\sigma = 0$
the Serret - Frenet equation (SFE).    Let us rewrite the equation (18)
in the form
$$
U_{t}-V_{x}+[U,V]=0  \eqno(19)
$$
where
$$
U =
\frac{1}{2i}\left ( \begin{array}{cc}
\tau        & k -i\sigma \\
k+i\sigma   & -\tau
\end{array} \right) ,\quad
V =
\frac{1}{2i}
\left ( \begin{array}{cc}
\omega_{1}             & \omega_{3}-i\omega_{2} \\
\omega_{3}+i\omega_{2} & - \omega_{1}
\end{array} \right) \eqno(20)
$$
The Lax representation (LR) of the equation (18)-(19) is given by
$$
\psi_{x} = U \psi \eqno(21a)
$$
$$
\psi_{t} =V\psi  \eqno(21b)
$$

Now let us give the interpretation of the above considered equations from
the point of view of our formalism. In our formalism, usually,
we assume that [8]
$$
k=\sum_{j} \lambda^{j}k_{j}, \quad
\tau=\sum_{j} \lambda^{j}\tau_{j}, \quad
\sigma=\sum_{j} \lambda^{j}\sigma_{j}, \quad
\omega_{k}=\sum_{j} \lambda^{j}\omega_{kj} \eqno(22a)
$$
where $j=0, \pm 1, \pm 2, \pm 3, ...$ .  Sometimes, instead of (22a) we take
the following more general case
$$
k=\sum_{j} h_{1j}k_{j}, \quad
\tau=\sum_{j} h_{2j}\tau_{j}, \quad
\sigma=\sum_{j} h_{3j}\sigma_{j}, \quad
\omega_{k}=\sum_{j} h_{4j}\omega_{kj} \eqno(22b)
$$
Here $h_{ij}= h_{ij}(\lambda)$ are some functions of $\lambda$ , where
   $\lambda$ is some characteristic
parameter of curves or surfaces or some  function of such parameters.
In the cases (22), i.e. when
$k, \tau, \sigma, \omega_{j}$ are some functions of $\lambda$,
the equations (15), (18)=(19) and  (21),  we call the (1+1)-dimensional
mM-LXI, mM-LXII and M-LXIX equations respectively.
[ These  conditional notations we use in order to accurately distinguish
these equations from the case when $k, \tau, \sigma, \omega_{j}$ are
independent of $\lambda$ and for convenience in our internal working
kitchen ].   Particular cases.

i) {\it The Gauss-Weingarden  and Gauss-Codazzi-Mainardi-Peterson equations}.
These equations  correspond to the case
$$
k=k_{0}, \quad
\tau=\tau_{0}, \quad
\sigma=\sigma_{0}, \quad
\omega_{k}=\omega_{k0} \eqno(23)
$$
i.e., when $k, \tau, \sigma $ are independent of $\lambda$.

ii) {\it The ZS-AKNS problem}.
The famous Zakharov-
Shabat-Ablowitz-Kaup-Newell-Segur (ZS-AKNS)  spectral
problem which generate many soliton equations in 1+1  is the
particular case of the M-LXIX equation (21a) as
$$
k_{0}=i(p+q), \quad k_{j}=0, \quad j\neq 0 \eqno(24a)
$$
$$
\sigma_{0} = p-q, \quad \sigma_{j} = 0, \quad j \neq 0  \eqno(24b)
$$
$$
\tau_{1} = -2, \quad \tau_{j} = 0, \quad j \neq 1  \eqno(24c)
$$

iii) {\it The Kaup-Newell-Wadati-Konno-Ichicawa spectral problem}.
This case corresponds to the reduction
$$
k_{1}=i (p+q), \quad k_{j}=0, \quad j\neq 1 \eqno(25a)
$$
$$
\sigma_{1} = p-q, \quad \sigma_{j} = 0, \quad j \neq 1  \eqno(25b)
$$
$$
\tau_{1} = -2, \quad \tau_{j} = 0, \quad j \neq 1  \eqno(25c)
$$

iv) {\it The  equations of a principal chiral field}.
The equations of a principal chiral field for functions $u, v$
$$
u_{t}+\frac{1}{2}[u,v]=0, \quad v_{x}-\frac{1}{2}[u,v]=0    \eqno(26)
$$
can equally be represented with the aid of (22b), if we choose
in this case
$$
U=\frac{u}{1-\lambda}, \quad V=\frac{v}{\lambda + 1}  \eqno(27)
$$

v) {\it The (1+1)-dimensional mM-LXVI equation}.
This case corresponds to the reduction when
$$
\tau^{2}\pm k^{2}\pm\sigma^{2} = n^{2} \eqno(28)
$$
The mM-LXVI equation contains in particular spin systems. In fact,
let
$$
k=nS_{1}, \quad \sigma = nS_{2}, \quad \tau=nS_{3}
\eqno(29)
$$
Then we get
$$
S_{3}^{2}\pm S_{1}^{2}\pm S_{2}^{2} = 1 \eqno(30)
$$
So that in this case the M-LXIX equation (21) is LR of the
(1+1)-dimensional M-O equation
$$
S_{t} -\frac{1}{n}V_{x}+[S, V]=0 \eqno(31)
$$
where
$$
S =
\left ( \begin{array}{cc}
S_{3}        & S^{-} \\
S^{+}   & -S_{3}
\end{array} \right) ,\quad
S^{\pm} = S_{1}\pm iS_{2} \eqno(32)
$$

So we see that practically all (1+1)-dimensional soliton equations
can be obtained from the mM-LXII equation (18)=(19) as some reductions.

\section{SDYM equation}
It is standard to define
a Yang-Mills vector  bundle over a four-dimensional manifold $M$
with connection one-form $A=A_{\mu}(x^{\nu})dx^{\mu}$.
The SDYM equations in this manifolds let us  write to set our notation:
$$
F = \ast F  \eqno(33)
$$
where $F$ is a curvature 2-form pulled back to M from the gauge bundle
$P(M, g)$, explicitly:
$$
F = d\omega + \omega \wedge \omega  \eqno(34)
$$
Here the connection 1-form $\omega$ on $P$ takes values in the Lie
algebra $g$ of the gauge group $G$. In terms of Cartesian coordinates
$x^{\mu}$, they can be expressed as
$$
F_{\mu\nu} = \frac{1}{2}\epsilon_{\mu\nu\kappa\sigma}F_{\kappa\sigma}
\eqno(35)
$$
where $\mu, \nu, ... =1,2,3,4,  \quad \epsilon_{\mu\nu\kappa\sigma}$ stands
for the completely antisymmetric
tensor in four dimensions with the convention: $\epsilon_{1234} = 1$.  The
components of the field strength $(F_{\mu\nu})$ are given by
$$
F_{\mu\nu} = [D_{\mu}, D_{\nu}] = \partial_{\mu}A_{\nu} - \partial_{\nu}A_{\mu}
+ [A_{\mu}, A_{\nu}]\eqno(36)
$$
where
$$
D_{\mu}=\partial_{\mu}-A_{\mu}
$$
are covariant derivatives.  We note that the SDYM equations (38) are invariant under the gauge transformation
$$
A_{\mu} \rightarrow \phi^{-1} A_{\mu} \phi - \phi^{-1}\phi_{\mu} \eqno(37)
$$

Let us introduce  the standard  null coordinates  in Euclidean space
$$
x_{\alpha}=t+iz, \quad x_{\bar \alpha}=t-iz, \quad
x_{\beta}=x+iy, \quad x_{\bar \beta}=x-iy
$$
In this coordinates the metric is given by
$$
ds^{2}=dx_{\alpha} dx_{\bar\alpha} + dx_{\beta}dx_{\bar\beta}
$$
Furthermore we have
$$
A_{\alpha} =A_{t}+iA_{z}, \quad A_{\bar \alpha}=A_{t}-iA_{z},
\quad A_{\beta}=A_{x}+iA_{y}, \quad A_{\bar\beta}=A_{x}-iA_{y}
$$
In this notations the SDYM equation are given by
$$
F_{\alpha\beta} = 0  \eqno(38a)
$$
$$
F_{\bar \alpha \bar \beta} = 0  \eqno(38b)
$$
$$
F_{\alpha\bar\alpha} - F_{\beta\bar\beta} = 0  \eqno(38c)
$$
or
$$
\partial_{\alpha}A_{\beta}-\partial_{\beta}A_{\alpha}
-[A_{\alpha}, A_{\beta}] = 0 \eqno(39a)
$$
$$
\partial_{\bar\alpha}A_{\bar\beta}-\partial_{\bar\beta}A_{\bar\alpha}
-[A_{\bar\alpha}, A_{\bar\beta}] = 0 \eqno(39b)
$$
$$
\partial_{\alpha}A_{\bar\alpha}-\partial_{\bar\alpha}A_{\alpha}
-[A_{\alpha}, A_{\bar\alpha}] =
\partial_{\beta}A_{\bar\beta}-\partial_{\bar\beta}A_{\beta}
-[A_{\beta}, A_{\bar\beta}] \eqno(39c)
$$
For the  equations (38)=(39), the LR has the form
$$
L\Phi = 0,  \quad M\Phi = 0  \eqno(40)
$$
where
$$
L = D_{\alpha} + \lambda D_{\bar\beta}, \quad M= D_{\beta}-
\lambda D_{\bar\alpha}  \eqno(41)
$$
and $\lambda$ is the spectral parameter.

\section{The (2+1)-dimensional mM-LXI and mM-LXII equations in space case}

In this section, we present some information from the (2+1)-dimensional
soliton geometry. One of the extension of the (1+1)-dimensional
M-LXI equation (15) is the following  (2+1)-dimensional mM-LXI equation
$$
\left ( \begin{array}{c}
{\bf e}_{1} \\
{\bf e}_{2} \\
{\bf e}_{3}
\end{array} \right)_{x}= A
\left ( \begin{array}{ccc}
{\bf e}_{1} \\
{\bf e}_{2} \\
{\bf e}_{3}
\end{array} \right) \eqno(42a)
$$
$$
\left ( \begin{array}{ccc}
{\bf e}_{1} \\
{\bf e}_{2} \\
{\bf e}_{3}
\end{array} \right)_{y}= B
\left ( \begin{array}{ccc}
{\bf e}_{1} \\
{\bf e}_{2} \\
{\bf e}_{3}
\end{array} \right) \eqno(42b)
$$
$$
\left ( \begin{array}{ccc}
{\bf e}_{1} \\
{\bf e}_{2} \\
{\bf e}_{3}
\end{array} \right)_{t}= D
\left ( \begin{array}{ccc}
{\bf e}_{1} \\
{\bf e}_{2} \\
{\bf e}_{3}
\end{array} \right)
\eqno(42c)
$$
where
$$
A =
\left ( \begin{array}{ccc}
0             & k     &-\sigma \\
-\beta k      & 0     & \tau  \\
\beta\sigma   & -\tau & 0
\end{array} \right) ,
\quad
B=
\left ( \begin{array}{ccc}
0            & m_{3}  & -m_{2} \\
-\beta m_{3} & 0      & m_{1} \\
\beta m_{2}  & -m_{1} & 0
\end{array} \right)
$$
$$
D=
\left ( \begin{array}{ccc}
0            & \omega_{3}  & -\omega_{2} \\
-\beta \omega_{3} & 0      & \omega_{1} \\
\beta \omega_{2}  & -\omega_{1} & 0
\end{array} \right).  \eqno(43)
$$

Hence, we obtain the following (2+1)-dimensional mM-LXII equation [8]
$$
A_{y} - B_{x} + [A,B] = 0 \eqno (44a)
$$
$$
A_{t} - D_{x} + [A,D] = 0 \eqno (44b)
$$
$$
B_{t} - D_{y} + [B,D] = 0 \eqno (44c)
$$

This equation admits the following LR
$$
g_{x} = Ug, \quad g_{y} = Vg, \quad  g_{t} = Tg \eqno(45)
$$

with
$$
U =
\frac{1}{2i}\left ( \begin{array}{cc}
\tau        & k -i\sigma \\
k+i\sigma   & -\tau
\end{array} \right) ,\quad
T =
\frac{1}{2i}
\left ( \begin{array}{cc}
\omega_{1}             & \omega_{3}-i\omega_{2} \\
\omega_{3}+i\omega_{2} & - \omega_{1}
\end{array} \right)
$$
$$
V =
\frac{1}{2i}\left ( \begin{array}{cc}
m_{1}        & m_{3} -im_{2} \\
m_{3}+im_{2}   & -m_{1}
\end{array} \right)
\eqno(46)
$$

So that the mM-LXII equation (44) we can rewrite in the following form
$$
F_{xy} = U_{y}-V_{x} + [U,V]=0  \eqno(47a)
$$
$$
F_{xt}=U_{t}-T_{x} + [U,T]=0  \eqno(47b)
$$
$$
F_{yt}=V_{t}-T_{y} + [V,T]=0  \eqno(47c)
$$
We note that the LR (45) we can write  in the more usual form
$$
Lg = 0, \quad Mg = 0 \eqno(48)
$$
where, for example, $L, M$ we can take in the form
$$
L = \partial_{x}+a\lambda \partial_{y}  -
(U+a\lambda V), \eqno(49a)
$$
$$
M=\partial_{t}+e\lambda^{2}\partial_{y}
-(T+e\lambda^{2}V) \eqno(49b)
$$
or
$$
L = D_{x}+a\lambda D_{y}, \quad M=D_{t}+e\lambda^{2}D_{y}. \eqno(50)
$$

Some (2+1)-dimensional soliton equations such as:
DS, Zakharov, (2+1)-complex mKdV, (2+1)-dNLS and so on, are exact
reductions of the mM-LXII equation (44)=(47). As example, in next
sections we show how obtain the well known representatives of (2+1)-SEs
- the DS and KP equations and their spin counterparts from the
mM-LXI and mM-LXII equations.

\subsection{Soliton equations in 2+1 dimensions as exact reductions
of the mM-LXII equation}

In this section, we obtain the IE and DS equation from
the mM-LXI and mM-LXII equations as some exact reductions.
The IE reads as [29]
$$
{\bf S}_{t}  =  {\bf S}\wedge ({\bf S}_{xx} +\alpha^2 {\bf S}_{yy})+
u_x{\bf S}_{y}+u_y{\bf S}_{x} \eqno (51a)
$$
$$
u_{xx}-\alpha^2 u_{yy}  = -2\alpha^2 {\bf S}\cdot ({\bf S}_{x}\wedge
                        {\bf S}_{y}).   \eqno (51b)
$$
We take the following identification
$$
{\bf S}={\bf e}_{1} \eqno(52)
$$
In this case we have
$$
m_{1}=\partial_{x}^{-1}[\tau_{y}-\frac{\beta}{2\alpha^2}M_2^{Ish}u] \eqno(53a)
$$
$$
m_{2}= -\frac{1}{2\alpha^2 k}M_2^{Ish}u \eqno (53b)
$$
$$
m_{3} =\partial_{x}^{-1}[k_y +\frac{\tau}{2\alpha^2 k}M_2^{Ish}u] \eqno(53c)
$$
$$
M^{IE}_{2}u=u_{xx}-\alpha^{2}u_{yy} \eqno(54)
$$
and
$$
\omega_{1} = \frac{1}{k}[-\omega_{2x}+\tau\omega_{3}]
                 \eqno (55a)
$$
$$
\omega_{2}= -k_{x}-
 \alpha^{2}(m_{3y}+m_{2}m_{1})+im_{2}u_{x}
\eqno (55b)
$$
$$
\omega_{3}= -k \tau+\alpha^{2}(m_{2y}-m_{3}m_{1})
+ik u_{y}+im_{3}u_{x}.
\eqno (55c)
$$
Now let us introduce two complex functions $q, p$, according to the formulae
$$
q = a_{1}e^{ib_{1}}, \quad p=a_{2}e^{ib_{2}} \eqno(56)
$$
Let $a_{j}, b_{j}$ have the forms
$$
a_{1}^2 =\frac{1}{4}k^2+
\frac{|\alpha|^2}{4}(m_3^2 +m_2^2)-\frac{1}{2}\alpha_{R}km_3-
\frac{1}{2}\alpha_{I}km_2
  \eqno(57a)
$$
$$
b_1 =\partial_{x}^{-1}\{-\frac{\gamma_1}{2ia_{1}^{2}}-
(\bar A-A+D-\bar D)\}  \eqno(57b)
$$
$$
a_2^2=\frac{1}{4}k^2+
\frac{|\alpha|^2}{4}(m_3^2 +m_2^2)+\frac{1}{2}\alpha_{R}km_3
-\frac{1}{2}\alpha_{I}km_2
  \eqno(57c)
$$
$$
b_{2} =\partial_{x}^{-1}\{-\frac{\gamma_2}{2ia_2^{2}}-
(A-\bar A+\bar D-D)\}  \eqno(57d)
$$
where
$$
\gamma_1=i\{\frac{1}{2}k^{2}\tau+
\frac{|\alpha|^2}{2}(m_3km_1+m_2k_y)-
$$
$$
\frac{1}{2}\alpha_{R}(k^{2}m_1+m_3k\tau+
m_2k_x)
+\frac{1}{2}\alpha_{I}[k(2k_y-m_{3x})-
k_x m_3]\}. \eqno(58a)
$$
$$
\gamma_2=-i\{\frac{1}{2}k^{2}\tau+
\frac{|\alpha|^2}{2}(m_3km_1+m_2 k_y)+
$$
$$
\frac{1}{2}\alpha_{R}(k^{2}m_1+m_3k\tau+
m_2k_x )
+\frac{1}{2}\alpha_{I}[k(2k_y-m_{3x})-
k_x m_3]\}. \eqno(58b)
$$
Here $\alpha=\alpha_{R}+i\alpha_{I}$. In this case, $q,p$ satisfy the following
DS  equation
$$
iq_t  + q_{xx}+\alpha^{2}q_{yy} + vq = 0 \eqno (59a)
$$
$$
-ip_t +  p_{xx}+\alpha^{2}p_{yy} + vp = 0 \eqno (59b)
$$
$$
v_{xx}-\alpha^{2}v_{yy} + 2[(p q)_{xx}+\alpha^{2}(p q)_{yy}] = 0.
\eqno (59c)
$$
It is means that  the IE (51) and the DS equation (59) are
L-equivalent [9] to each other.
As well known that these equations are G-equivalent [22] to each other [30].
A few comments are in order.

i) From these results, we get the Ishimori I and DS-I equations as
$\alpha_{R}=1, \alpha_{I}=0$

ii) and the Ishimori II and DS-II equations as $\alpha_{R}=0, \alpha_{I}=1$.

iii) For  DS-II equation we have
$$
pq=\mid q\mid^{2}=\mid p\mid^{2}  \eqno(60)
$$

iv) at the same time, for the DS-I equation we obtain
$$
pq\neq\mid q\mid^{2}\neq\mid p\mid^{2}  \eqno(61)
$$
$$
\mid q\mid^{2}=\mid p\mid^{2}-km_{3}  \eqno(62)
$$
$$
pq= (pq)_{R}+i(pq)_{I}
\eqno(63)
$$
so that  $pq$ is the complex quantity.

\section{The mM-LXI and mM-LXII equations in the plane case}

The (2+1)-dimensional mM-LXI equation in plane has the form [8]
$$
\left ( \begin{array}{cc}
{\bf e}_{1} \\
{\bf e}_{2}
\end{array} \right)_{x}= A_{p}
\left ( \begin{array}{cc}
{\bf e}_{1} \\
{\bf e}_{2}
\end{array} \right), \quad
\left ( \begin{array}{cc}
{\bf e}_{1} \\
{\bf e}_{2}
\end{array} \right)_{y}= B_{p}
\left ( \begin{array}{cc}
{\bf e}_{1} \\
{\bf e}_{2}
\end{array} \right), \quad
\left ( \begin{array}{cc}
{\bf e}_{1} \\
{\bf e}_{2}
\end{array} \right)_{t}= D_{p}
\left ( \begin{array}{cc}
{\bf e}_{1} \\
{\bf e}_{2}
\end{array} \right)
\eqno(64)
$$
where
$$
A_{p} =
\left ( \begin{array}{cc}
0             & k     \\
-\beta k      & 0
\end{array} \right) ,
\quad
B_{p}=
\left ( \begin{array}{cc}
0            & m_{3}   \\
-\beta m_{3} & 0
\end{array} \right)
$$
$$
D_{p}=
\left ( \begin{array}{cc}
0            & \omega_{3}   \\
-\beta \omega_{3} & 0
\end{array} \right).  \eqno(65)
$$

In the plane case the mM-LXII equation takes the following simple form
$$
k_{y} = m_{3x}   \eqno(66a)
$$
$$
k_{t} = \omega_{3x} \eqno(66b)
$$
$$
m_{3t}=\omega_{3y}  \eqno(66c)
$$
Hence we get
$$
m_{3} = \partial^{-1}_{x}k_{y}\eqno (67)
$$
The nonlinear evolution equation has the form  (66b).
Many (2+1)-dimensional integrable equations such as the
Kadomtsev-Petviashvili, Novikov-Veselov (NV), mNV, KNV, (2+1)-KdV, mKdV
equations are the integrable reductions of
the M-LXII equation (66).

\subsection{The KP and mKP equations as exact reductions of the mM-LXII equation}

For example, let us show that the KP and  mKP equations are exact
reductions of the mM-LXII equation (66).
Consider the M-X equation [8]
$$
{\bf S}_t = \frac{\omega_{3}}{k}{\bf S}_{x}  \eqno(68)
$$
where
$$
\omega_{3}= -k_{xx}-3k^{2}--3\alpha^{2}\partial^{-1}_{x}m_{3y} \eqno(69)
$$
If we put ${\bf S}= {\bf e}_{1}$ then  from (66) we obtain the L-equivalent
counterpart of the M-X equation which is the KP equation
$$
k_{t} +6kk_{x} + k_{xxx} +3\alpha^{2}m_{3y}=0 \eqno(70a)
$$
$$
m_{3x}=k_{y} \eqno(70b)
$$
As known the LR of this equation is given by
$$
\alpha\psi_{y}+\psi_{xx}+k\psi = 0 \eqno(71a)
$$
$$
\psi_{t}+4\psi_{xxx}+6k\psi_{x}+3(k_{x}-\alpha m_{3})\psi =0 \eqno(71b)
$$
\section{Linear problems from the mM-LXII equation}

The idea is the following. Let us rewrite the mM-LXII equation (44)
in the form
$$
LA=B_{x} \eqno(72a)
$$
$$
MA=D_{x} \eqno(72b)
$$
where
$$
L=\partial_{y}-[\dot, B], \quad M=\partial_{t}-[\dot, D] \eqno(73)
$$
For the case (47) these equations take the forms
$$
LU=V_{x} \eqno(74a)
$$
$$
MU=T_{x} \eqno(74b)
$$
The compatibility condition of the equations (72) and (74) are the
equations (44c) and (47c), respectively. So that the equations (72) and (74)
play the role of the LR for the evolution equations (44c) and (47c),
respectively.
In this way, we can obtain new SEs. Example. For simplicity, we consider
the M-LXII equation (66). The compatibility condition of the equations
(66a,b) is the equation (66c). Let   we choose
$$
m_{3}= -\frac{1}{\alpha}(k_{x}+k^{2}+u), \quad
\omega_{3}=-4k_{xx}-12kk_{x}-4k^{3}-6uk-3u_{x}+3\alpha v \eqno(75)
$$
Then the compatibility condition of (66a,b) gives
$$
u_{t} +6uu_{x} + u_{xxx} +3\alpha^{2}v_{y}=0 \eqno(76a)
$$
$$
v_{x}=u_{y} \eqno(76b)
$$
which is nothing but the KP equation (70). On the other hand, eliminating
$u$ out of equations (66a,b) with (75), one obtains the modified KP (mKP)
equation
$$
k_{t}= 6k^{2}k_{x}-k_{xxx}+3\alpha(2k_{x}w-\alpha w_{y})  \eqno(77a)
$$
$$
w_{x}=k_{y} \eqno(77b)
$$
It is interesting to note that if define $k$ by
$$
k= \frac{\psi_{x}}{\psi} \eqno(78)
$$
then $\psi$ satisfies the equations (71). Finally we note that the
(2+1)-dimensional M-LXI and M-LXII equations are the particular cases
of the (2+1)-dimensional mM-LXI
and mM-LXII equations as $\sigma = 0$ respectively.

\section{Soliton equations in 2+1 as reductions of the SDYM equation}

As in introduction mentioned the SDYM equation contains the (1+1)-dimensional
SEs as particular reductions. In this section we show that SEs in 2+1
dimensions also are exact reductions of the SDYM equation (39).
For this purpose, consider the coordinates
$$
x_{\alpha} = it, \quad x_{\bar\alpha} = -it, \quad x_{\beta} = x+iy, \quad
x_{\bar\beta} = x-iy \eqno(79)
$$
Now in the SDYM equation (38)=(39) we  take
$$
A_{\alpha}=-iD, \quad A_{\bar\alpha}=iD, \quad A_{\beta}=A-iB,
\quad A_{\bar\beta}=A+iB.  \eqno(80)
$$
where we mention that $A,B,C,D$  are in   our case real matrices.
Then the SDYM equation (39) reduces to the (2+1)-dimensional
mM-LXII equation (44).
So the (2+1)-dimensional mM-LXII equation is the integrable reduction
of the SDYM equation.

As many (may be all) soliton equations in 2+1 dimensions
are some integrable reductions of the mM-LXII and/or M-LXII equations
(44) and/or (47)  then as follows
from the results of the previous sections these (2+1)-dimensional
soliton equations are exact reductions of the SDYM equation.

\section{The Bogomolny equation and SEs }

It is well known that if  in the SDYM equation (39) we take
$$
A_{\alpha}=\Psi-iD, \quad A_{\bar\alpha}=\Psi+iD, \quad A_{\beta}=A-iB,
\quad A_{\bar\beta}=A+iB.  \eqno(81)
$$
and assume that $A, B, D, \Psi$ are independent of $z$, then we obtain
the following  equation
$$
\Psi_{t}+[\Psi, D]+A_{y} - B_{x} + [A,B] = 0 \eqno (82a)
$$
$$
\Psi_{y}+[\Psi, B]+D_{x}-A_{t}  + [D, A] = 0 \eqno (82b)
$$
$$
\Psi_{x}+[\Psi, A]+B_{t} - D_{y} + [B,D] = 0 \eqno (82c)
$$
which is nothing but  the  Bogomolny equation (BE) in Euclidean coordinates, which
as known is relevant in the study of magnetic monopoles [31-34]. If in the
BE (82) we put $\Psi = 0$, then we get the mM-LXII equation (44). So we have
also shown that the mM-LXII equation and hence SEs in 2+1 dimensions are
the particular exact reductions of the BE.

\section{The M-LXVI  equation  and spin systems}

In this section we show how spin systems can be included into our formalism.
For this purpose, we consider the M-LXVI equation which is the particular
case of some above considered equations as
$$
\tau^{2} \pm k^{2}\pm \sigma^{2} =  n^{2}(x,y,t)  \eqno(83)
$$
Let
$$
k = n S_{1}, \quad \sigma = n S_{2}, \quad \tau = n S_{3}  \eqno(84)
$$
Then from (83) follows that
$$
S_{3}^{2} \pm S_{1}^{2} \pm S_{2}^{2} = 1 \eqno(85)
$$
Below we consider the case when $\tau^{2}+k^{2}+\sigma^{2}=n^{2}=constant$.
Let us we deduce the (2+1)-dimensional M-LXVI equation. For this purpose,
we consider the gauge transformation
$$
\left ( \begin{array}{ccc}
{\bf f}_{1} \\
{\bf f}_{2} \\
{\bf f}_{3}
\end{array} \right)= E
\left ( \begin{array}{ccc}
{\bf e}_{1} \\
{\bf e}_{2} \\
{\bf e}_{3}
\end{array} \right)
\eqno(86)
$$
where ${\bf e}_{j}$ are the solutions of the mM-LXI equation (42).
Then hence and from (42)  we get
$$
\left ( \begin{array}{ccc}
{\bf f}_{1} \\
{\bf f}_{2} \\
{\bf f}_{3}
\end{array} \right)_{x}= A^{\prime}
\left ( \begin{array}{ccc}
{\bf f}_{1} \\
{\bf f}_{2} \\
{\bf f}_{3}
\end{array} \right) \eqno(87a)
$$
$$
\left ( \begin{array}{ccc}
{\bf f}_{1} \\
{\bf f}_{2} \\
{\bf f}_{3}
\end{array} \right)_{y}= B^{\prime}
\left ( \begin{array}{ccc}
{\bf f}_{1} \\
{\bf f}_{2} \\
{\bf f}_{3}
\end{array} \right) \eqno(87b)
$$
$$
\left ( \begin{array}{ccc}
{\bf f}_{1} \\
{\bf f}_{2} \\
{\bf f}_{3}
\end{array} \right)_{t}= D^{\prime}
\left ( \begin{array}{ccc}
{\bf f}_{1} \\
{\bf f}_{2} \\
{\bf f}_{3}
\end{array} \right) \eqno(87c)
$$
Here
$$
A^{\prime}=EAE^{-1}+E_{x}E^{-1}, \quad
B^{\prime}=EBE^{-1}+E_{x}E^{-1}, \quad
D^{\prime}=EDE^{-1}+E_{x}E^{-1} \eqno(88)
$$
We can choose the function $E$ so that
$$
A^{\prime} =a
\left ( \begin{array}{ccc}
0             & S_{1}     & -S_{2} \\
-\beta S_{1}      & 0     & S_{3}  \\
\beta S _{2}  & -S_{3} & 0
\end{array} \right)
\eqno(89)
$$
Then the compatibility condition of the equations (87) gives
the (2+1)-dimensional M-0 equation
$$
A^{\prime}_{y} - B^{\prime}_{x} + [A^{\prime}, B^{\prime}] = 0 \eqno (90a)
$$
$$
A^{\prime}_{t} - D^{\prime}_{x} + [A^{\prime}, D^{\prime}] = 0 \eqno (90b)
$$
$$
B^{\prime}_{t} - D^{\prime}_{y} + [B^{\prime}, D^{\prime}] = 0 \eqno (90c)
$$
The LR of this equation has the form
$$
\psi_{x} = aS \psi \eqno(91a)
$$
$$
\psi_{y} =V^{\prime}\psi  \eqno(91b)
$$
$$
\psi_{t} =T^{\prime}\psi  \eqno(91c)
$$
So that the (2+1)-dimensional M-0 equation take the form
$$
S_{t}-\frac{1}{a}V^{\prime}_{x}+[S,V^{\prime}]=0  \eqno(92a)
$$
$$
S_{y}-\frac{1}{a}T^{\prime}_{x}+[S,T^{\prime}]=0  \eqno(92b)
$$
$$
V^{\prime}_{t}-T^{\prime}_{y}+[V^{\prime}, T^{\prime}]=0  \eqno(92c)
$$

Finally we note that the M-0 equation (90)=(92) admits many integrable spin
systems in 2+1 dimensions.

\section{Summary}

So,  we have considered some aspects of the multidimensional
soliton geometry. Our approach permits find some integrable classess
of curves and surfaces in multidimensions. Also we have shown that the
mM-LXII equation is exact reduction of the Bogomolny hence and SDYM
equations. As  many soliton equations ,
for example, in 2+1 dimensions are particular cases of the mM-LXII
(and/or M-LXII) equation, it is means that  they are in turn exact reductions
of the SDYM equation and/or the Bogomolny equation.
The connection between  spin systems and
curves/surfaces  is also discussed.  It is shown that the M-0 equation
which generate spin systems also is exact reduction of the Bogomolny
and SDYM equations.   So the Ward's conjecture (see Introduction) is
justified (once more) and in 2+1 dimensions. However, many
questions remain open and deserve further investigation. So the further
studies of these questions seem to be very interesting.

A few words on soliton geometry in 3+1 dimensions. In 3+1 dimensions, the
equation (15) admits several extensions. Some of them are as follows:

i) The (3+1)-dimensional mM-LXI equation [8]
$$
\left ( \begin{array}{c}
{\bf e}_{1} \\
{\bf e}_{2} \\
{\bf e}_{3}\\
\vdots \\
{\bf e}_{n}
\end{array} \right)_{x}= A
\left ( \begin{array}{c}
{\bf e}_{1} \\
{\bf e}_{2} \\
{\bf e}_{3} \\
\vdots  \\
{\bf e}_{n}
\end{array} \right), \quad
\left ( \begin{array}{c}
{\bf e}_{1} \\
{\bf e}_{2} \\
{\bf e}_{3}\\
\vdots \\
{\bf e}_{n}
\end{array} \right)_{y}= B
\left ( \begin{array}{c}
{\bf e}_{1} \\
{\bf e}_{2} \\
{\bf e}_{3} \\
\vdots  \\
{\bf e}_{n}
\end{array} \right)
$$
$$
\left ( \begin{array}{c}
{\bf e}_{1} \\
{\bf e}_{2} \\
{\bf e}_{3}\\
\vdots \\
{\bf e}_{n}
\end{array} \right)_{z}= C
\left ( \begin{array}{c}
{\bf e}_{1} \\
{\bf e}_{2} \\
{\bf e}_{3} \\
\vdots  \\
{\bf e}_{n}
\end{array} \right), \quad
\left ( \begin{array}{c}
{\bf e}_{1} \\
{\bf e}_{2} \\
{\bf e}_{3}\\
\vdots \\
{\bf e}_{n}
\end{array} \right)_{t}= D
\left ( \begin{array}{c}
{\bf e}_{1} \\
{\bf e}_{2} \\
{\bf e}_{3} \\
\vdots  \\
{\bf e}_{n}
\end{array} \right)
\eqno(93)
$$

ii) The (3+1)-dimensional mM-LXVIII equation [8]
$$
\left ( \begin{array}{c}
{\bf e}_{1} \\
{\bf e}_{2} \\
{\bf e}_{3}\\
\vdots \\
{\bf e}_{n}
\end{array} \right)_{x_{\alpha}}= -\lambda
\left ( \begin{array}{c}
{\bf e}_{1} \\
{\bf e}_{2} \\
{\bf e}_{3} \\
\vdots  \\
{\bf e}_{n}
\end{array} \right)_{x_{\bar\beta}} + A
\left ( \begin{array}{c}
{\bf e}_{1} \\
{\bf e}_{2} \\
{\bf e}_{3}\\
\vdots \\
{\bf e}_{n}
\end{array} \right)
\eqno(94a)
$$
$$
\left ( \begin{array}{c}
{\bf e}_{1} \\
{\bf e}_{2} \\
{\bf e}_{3} \\
\vdots  \\
{\bf e}_{n}
\end{array} \right)_{x_{\beta}}=\lambda
\left ( \begin{array}{c}
{\bf e}_{1} \\
{\bf e}_{2} \\
{\bf e}_{3}\\
\vdots \\
{\bf e}_{n}
\end{array} \right)_{x_{\bar\alpha}}+B
\left ( \begin{array}{c}
{\bf e}_{1} \\
{\bf e}_{2} \\
{\bf e}_{3} \\
\vdots  \\
{\bf e}_{n}
\end{array} \right)
\eqno(94b)
$$
and so on.

\section{Acknowledgements}
RM would like to thank Prof. M. Lakshmanan for  hospitality
during  visits,  for many stimulating discussions and for the
financial support (that is extremely important personally
for me and for my group).
RM also would like to thank Radha Balakrishnan and M. Daniel for hospitality
and for useful discussions.

\section{Questions}
{\bf Guestion-1}: Please, give the twistor description of the above presented
results.        \\
{\bf Question-2}: Please, consider the connection between the above results
and the Self-Dual Einstein equation and hyper-Kahler  hierarchies.

\end{document}